\RequirePackage{fix-cm}
\documentclass[smallextended]{svjour3}       
\smartqed  
\usepackage[utf8]{inputenc}
\usepackage{graphicx}
\usepackage{mathrsfs}
\usepackage{amsbsy}
\usepackage{amsmath}
\usepackage{amssymb}
\begin{document}

\title{Single-phase flow through a porous media revisited}

\author{F. Mandujano         \and
        C. Málaga 
}


\institute{F. Mandujano \at
              Physics Department, School of Science, Universidad Nacional Aunt\'onoma de M\'exico.\\
              \email{frmas@ciencias.unam.mx}
           \and
           C. Málaga \at
            Physics Department, School of Science, Universidad Nacional Aunt\'onoma de M\'exico.\\
}

\date{Received: date / Accepted: date}

\maketitle

\begin{abstract}
The problem of the flow trough a porous media is formulated in terms of a pressure equation, based on arguments of volume conservation which state the mechanical equilibrium between the solid and the fluid phases. In the resulting governing equations, the conservation of mass is replaced by a transport-diffusion equation for the pressure field. When the deformation of the solid matrix is negligible, the pressure equation can be approximated by a non-linear diffusion equation. The behavior of the system of equations is tested under an oscillatory pressure gradient.
\end{abstract}

\section{Introduction}
\label{intro}
The study of the flow of a simple fluid, characterized by three state variables, trough a porous media consist of solving the conservation of mass and the balance of momentum on a Representative Element Volume (REV)~\cite{bear,chen,xu,kieu}. The conservation of mass takes the same form as in the case of the flow of a fluid, and there are several models that account for the balance of momentum~\cite{whitaker,ruth,fourar,cmei}. 

The simplest assumption, for a stationary flow, takes the Darcy's law~\cite{darcy} in place of the balance of momentum equation. For incompressible flows, the fluid density and the solid matrix porosity are assumed constant, and the problem is reduced to a Laplace equation for the pressure field~\cite{bear}.

On the contrary, when the flow is assumed compressible, the conservation principles are closed with an equation of state between the pressure and the fluid density such that the temporal variations of the pressure field are proportional to the mass flux~\cite{chen,kieu,whitaker,acs}. When the density is eliminated from the system of equations a transport-diffusion equation is found, that depends on the compresibilities of the fluid and the solid matrix~\cite{kieu,whitaker}. Assuming substances with very small compresibilities the equation for the pressure field results in a non-linear diffusion equation. This scheme is valid for flows at low values of the Reynolds number, when inertial effects can be neglected~\cite{bear}. Time dependent phenomena can be modeled using a modification of the Darcy's law that includes time dependent terms representing the inertia of the fluid~\cite{bernardi,zhu2016}. 

In the present work, we start with the assumption that the volume of the pores is fully occupied by a single phase fluid~\cite{acs}, a consequence of the mechanical equilibrium between the solid matrix and the fluid. Identifying the volume occupied by the fluid in a REV with the material volume, we arrive to a relation between the material derivative of the pressure field and the logarithmic material derivative of the number of fluid particles within the REV. The article is divided as follows; in section \ref{opf} the deduction of the pressure equation and some of its general properties are presented, in section \ref{samples} numerical examples for full time dependent numerical solutions are analyzed  and some conclusions are drawn in section \ref{con}.

\section{One phase flow}
\label{opf}
The study of the flow trough a porous media can be made assuming conservation principles on a Representative Element of Volume (REV) of the porous media~\cite{bear}. If $\xi$ is the number of fluid particles per unit volume and $\phi$ the porosity, the conservation of mass reads
\begin{equation}
 \frac{D n }{D t} +  n \nabla \cdot {\bf v} = q,\label{mass}
\end{equation}
were $n=\phi \xi$ is the number of fluid particles within the REV, ${\bf v}$ is the intrinsic velocity, $q$ a source of mass and \(\frac{D}{D t} = \frac{\partial}{\partial t} + {\bf v} \cdot \nabla \) is the material derivative. 

In view of the complexity of the problem, the Darcy's law is commonly used in place of the momentum balance equation, which can be viewed as a balance between viscous and pressure forces. Hence, if ${\bf u}$ represents the flux per unit volume that flows through the REV due to a pressure gradient the Darcy law is given by
\begin{equation}
{\bf u} = -\frac{\mathbb{K}}{\eta} \cdot (\nabla P - \rho {\bf g}),\label{darcy}
\end{equation}
where $\mathbb{K}$ is the permeability tensor of the porous media, $\eta$ the fluid dynamic viscosity, ${\bf g}$ the acceleration of gravity, $\rho$ is the fluid mass density and ${\bf u}= \phi {\bf v}$ is the relation between {\bf u} and the intrinsic velocity. Given the state equations as two functions of $P$, expressions (\ref{mass}) and (\ref{darcy}) result in a non linear transport equation for the pressure field. This scheme has been commonly used in the past~\cite{celik,medi,chen,birdsell}.

More recently, another point of view was proposed by Acs et. al.~\cite{acs} and is based in that the volume of the pores has to be equal to the volume of fluid within them at any time. The volume occupied by the fluid and the solid matrix are derivatives of the Gibbs free energy of the composite system within the REV, which for an isothermal process has the form $G_i=G_i(P_i,N_i)$ where $P_i$ is the pressure, $N_i$ the number of particles and the sub-index $i$ indicates the fluid or the solid matrix. 

Assuming local equilibrium within the REV the pressure and the volumes of the fluid and the solid matrix are the same, in view of the boundary condition of an elastic solid, therefore the partial derivatives of the free energies with respect to pressure are equal $V_p=V_f$, where $V_p=V_p(P)$ and $V_f=V_f(P,N)$ are the volume of the pore and the volume of fluid respectively and $N$ the number of particles of the fluid phase. The conservation of volumes can be stated conveniently as follows,
\begin{equation}
 \frac{1}{V_p}\frac{DV_p}{Dt} = \frac{1}{V_f}\frac{DV_f}{Dt},\label{vols}
\end{equation}
the left hand side of the above equation represents the material variation of the volume of the solid, as will be shown below. 

It has to be recalled that in Ref.~\cite{acs} equation (\ref{vols}) is stated with the partial derivative instead of the material derivative. However to be consistent with the definition of a thermodynamic volume, this corresponds to the Lagrangian volume element. When the derivatives of the volumes are written in terms of derivatives of the pressure field and the number of particles, it is found that
\begin{equation}
 \frac{1}{V_p} \frac{\partial V_p}{\partial P} \frac{D_s P }{D t} = \frac{1}{V_f} \frac{\partial V_f}{\partial P} \frac{D_s P }{D t} + \frac{1}{V_f} \frac{\partial V_f}{\partial N} \frac{D_f N }{D t}, 
\end{equation} 
where a sub-index is included to indicate that each phase, solid (s) or fluid (f), is transported with its intrinsic velocity in analogy with the treatment of a multiphase flow~\cite{bear}. The changes in pressure within the REV are due to the deformation of the solid matrix, hence the material derivatives for the pressure field are taken with the velocity of the solid phase ${\bf v}_s$. 

Given that $\phi = V_f/V_o$, $n=N/V_o$, where $V_o$ is the volume of the REV, and \[\frac{\partial V_f}{\partial N}=\frac{1}{\xi},\] the equality between volumes becomes
\begin{equation}
\left(\frac{1}{V_p}\frac{\partial V_p}{\partial P}-\frac{1}{V_f} \frac{\partial V_f}{\partial P}\right)\frac{D_s P}{Dt} = \frac{1}{n} \frac{D_f n }{D t},\label{pres1}
\end{equation}
hence a relation between the material derivatives of the pressure field and the density of particles within the REV is obtained.  

The partial derivatives inside the parenthesis can be written in terms of the porosity and the compressibilities of both the solid and the fluid phases. Using that $V_p=V_{o}-V_s$, where $V_s$ is the volume of the solid phase, they can be written as
\begin{eqnarray}
 \frac{1}{V_p}\frac{\partial V_p}{\partial P} &=& \left(\frac{1}{\phi}-1\right) C_s, \label{coms}\\
 \frac{1}{V_f}\frac{\partial V_f}{\partial P} &=&  -C_f,\label{comf}
\end{eqnarray}
where $C_s$ y $C_f$ are the compressibilities of the solid and fluid phases, respectively. Combining the above expressions with equation (\ref{mass}) we find
\begin{equation}
\left[ (1-\phi)C_r+\phi C_f \right] \frac{D_s P}{Dt} = -\phi \nabla \cdot {\bf v} + \frac{q\phi}{n},\label{pres2}
\end{equation}
which states that the temporal changes in the pressure field are due to the local compressions and expansions of fluid elements. This is different from previous proposals of the pressure equation~\cite{whitaker,celik,birdsell} where the temporal changes in the pressure field are due to the mass flux, instead we relate the pressure variations with the rate of change of the volume of the fluid element.

This equation has in principle the same information as that contained in equation (\ref{mass}) and may be taken as a dynamical
extension of the thermodynamic definition of compressibility of an effective medium given by
\begin{equation}
\mathcal{C}^{-1}(\phi,P) = \frac{\phi}{(1-\phi)C_s+\phi C_f},\nonumber 
\end{equation} 
in analogy with expressions (\ref{coms}) and (\ref{comf}).

To solve the pressure equation, information about the dynamics of the solid phase must be provided besides the equation of motion of the fluid phase. Hence, within this point of view the problem consist of two equations for the conservation of momentum, one for each phase, coupled at the solid-fluid boundary through equation (\ref{pres2}).

In many practical applications, the deformation of the solid matrix is negligible with respect to the intrinsic fluid velocities, therefore the material derivative at the left-hand side of equation (\ref{pres2}) is approximately equal to the partial time derivative. Using the Darcy's law in place of the balance of momentum equation, the evolution of the pressure field of a fluid flow through a porous media for small deformations of the solid matrix is given by
\begin{equation}
\frac{\partial P}{\partial t} = \frac{1}{\mathcal{C}(\phi,P)}\nabla \cdot \left(\frac{\mathbb{K}}{\eta\phi}\cdot( \nabla P - \rho \bf g )\right), \label{pres3}
\end{equation}
even when this assumption is also made by D. T. Birdsell et. al. in Ref.~\cite{birdsell}, their resulting equation for the pressure field relate its temporal variations with the fluid mass flux and the associated coefficient cannot be identified with a compressibility.   

When the pressure field is independent of time, it satisfies a Poisson equation, which corresponds to an incompressible uniform flow~\cite{bear}. Given the law of evolution of the velocity field ${\bf v}({\bf r},t)$ and two equations of state equation (\ref{pres3}) gives the evolution of the pressure field. 

When the permeability, the porosity and the fluid density are assumed constant, the resulting equation is a diffusion equation with a coefficient that depends on $\phi$ and the compressibilities. Three different limits can be identify depending on the relative value of $C_f$ and $C_s$. When $C_f/C_s \sim 1 $ the diffusion coefficient in (\ref{pres4}) is such that $\mathcal{C} \sim {\mathbb{K}}$. If the compressibility of the solid is much bigger than the fluid compressibility $\mathcal{C} \sim (1-\phi)\mathbb{K}$. On the contrary, if $C_f/C_s \gg 1$ then $\mathcal{C} \sim (1+\phi)\mathbb{K}$, and the diffusion coefficient is the highest when the fluid density changes become important. 

In a more general case, the pressure field changes also due to gradients of the porosity, in the case of a deformable solid matrix, and when the porous media is heterogeneous and anisotropic. For non-Darcyan flows, a momentum equation must be provided and in general has to be solved simultaneously with the pressure equation (\ref{pres2}).

\section{Oscillating flows}
\label{samples}
In order to present examples of the present scheme the oscillating flow of a rectangular region of length $L$, height $H$ and width $W$ filled with a porous media is simulated using finite differences in parallel on GPU's; the flow is produced by an oscillatory pressure difference $\Delta P$ at he boundaries in the $x$-direction of the form \[ \frac{ \Delta P}{L}= \frac{1}{L} \left(\Delta P_0+\delta \mathcal{P} \sin (\omega t)  \right),\] where $\Delta P_0$ and $\delta \mathcal{P}$ are constant and $\omega$ is the frequency of oscillation; the remaining walls, in the $y$ and $z$ directions, are assumed impermeable. 

We assume that the solid has a constant compressibility so that all the quantities and equations were scaled using $C_s$, the dynamic viscosity $\eta$ of the fluid and $L$. The partial differential equation for the pressure in dimensionless form is given by
\begin{equation}
\frac{\partial P}{\partial t} = \frac{\phi}{((1-\phi)+\phi C)}  \nabla \cdot \left(\frac{\mathbb{K}}{\phi}\cdot( \nabla P + \rho \hat{\bf k} )\right),\label{pres4}
\end{equation}
where $C=C_f/C_s$, the characteristic time and density are $\eta C_s$ and $g C_s L$ and it is assumed that there is any source within the domain; the permeability tensor is scaled with $L^2$. In the case when the fluid density, the porosity and the hydraulic permeability are constants, the stationary solution of equation (\ref{pres4}) in a closed system is the hydrostatic pressure with the fluid at rest as is expected.

When both compressibilities are constant, equations (\ref{coms}) and (\ref{comf}) can be integrated giving exponential laws that depends on a reference thermodynamic state. In the case of the solid
\begin{equation}
\phi = 1-(1-\phi_0)e^{-(P-P_0^s)},\label{phi}
\end{equation}
which is such that $ 0< \phi \leq 1$. Depending on the reference state, it can be possible to model situations where the changes in the porosity become important. This state equation is also consistent with the case of a non deformable solid with constant compressibility with an appropriate election of the reference state. 

A liquid with constant compressibility behaves similarly, but the state equation depends in general on a different reference state, the liquid density has the form
\begin{equation}
\rho = \rho_0 e^{C(P-P_0^f)},\label{rho}
\end{equation}
hence the fluid density increase as $C$ is increased. The assumed scaling is useful to go from a compressible fluid to an incompressible fluid; the porosity given by expression (\ref{phi}) is approximately constant for small deviations from the solid reference state.

A set of simulations were performed for different values of $\omega$ and $C$, the reference pressure of both substances are taken equal for simplicity and $\Delta P_0$ and $\delta \mathcal{P}$ are fixed. As initial condition, the pressure inside the medium was taken uniform and the numerical method is iterated until the flow became stationary or periodic.

\begin{figure}[h]
\begin{center}
 \includegraphics[width=0.9\columnwidth]{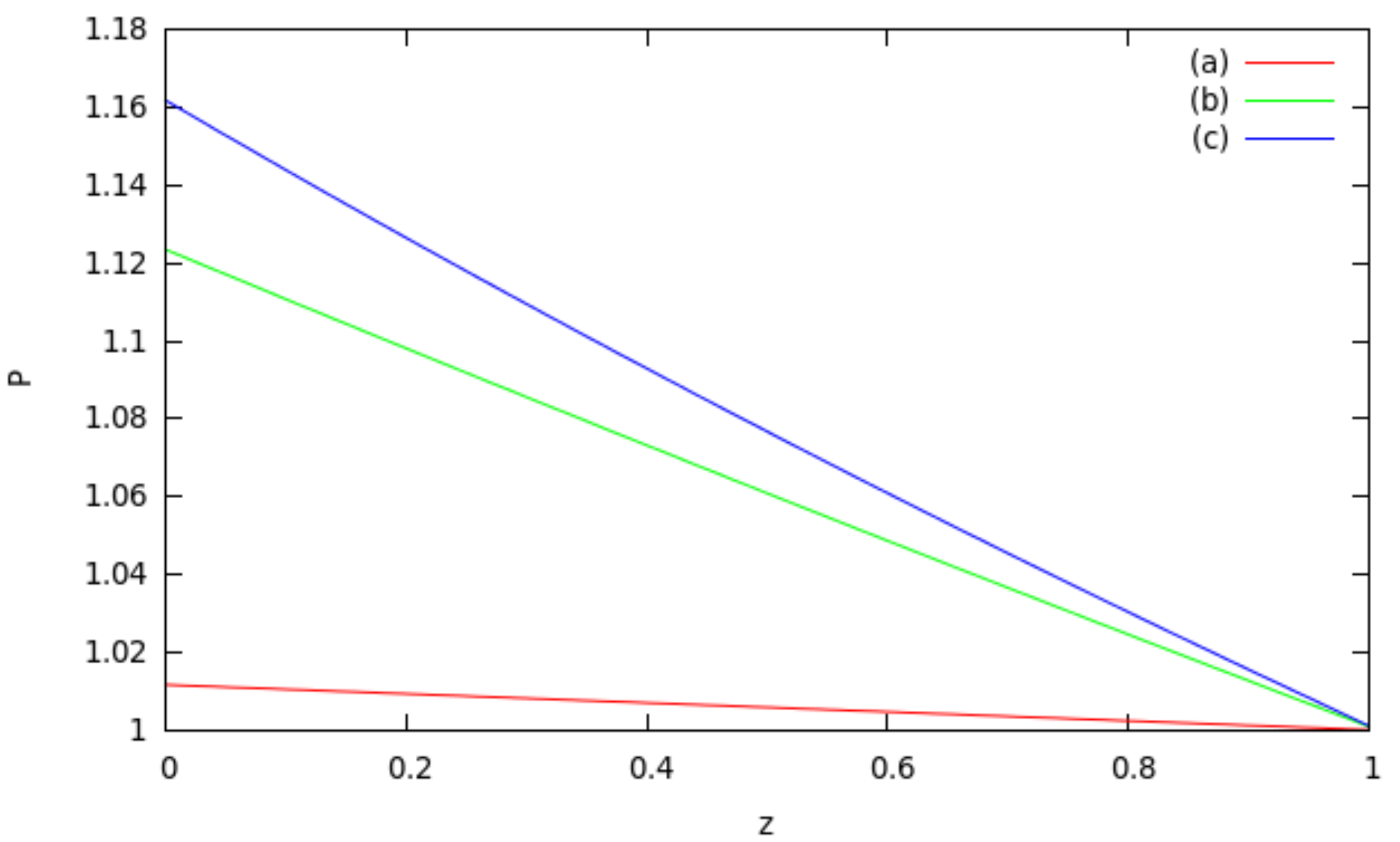} 
\end{center}
\caption{Plot of the pressure as function of $z$, the pressure at the top is $P_a=1$. a) $\rho_0 =0.01$ y $C=0.7$, b) $\rho_0=0.1$ y $C=0.7$. c) $\rho_0=0.1$ y $C=1.5$.} \label{fig:hid}
\end{figure}

When $\Delta P_0=0$ and $\delta \mathcal{P}=0$ the flow evolves from the uniform pressure field  and becomes stationary. For small values of $C$ and working near to the reference state, the pressure field becomes hydrostatic which corresponds to the stationary solution of equation (\ref{pres4}) with constant coefficients. As $C$ is increased, the solution starts to deviate from the hydrostatic solution as shown in figure \ref{fig:hid}.
The pressure at the bottom is increased due to the changes in the fluid density and the porosity. 

\begin{figure}[h]
\begin{center}
 \includegraphics[width=1.0\columnwidth]{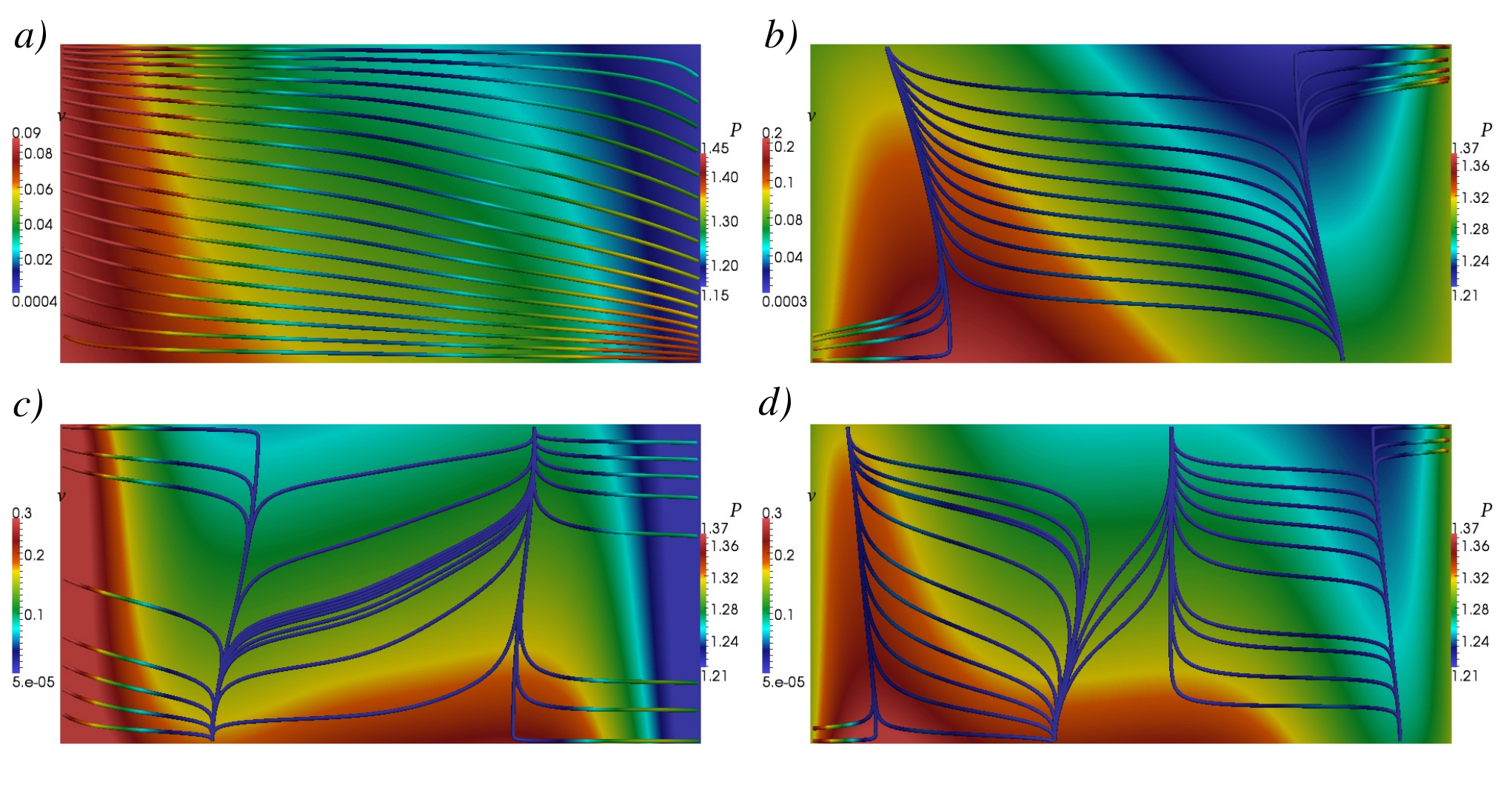} 
\end{center}
\caption{Pressure and stream lines in the plane $y=0$ for $C=0.1$ a) $t=0$ and b) $t=\pi/(2\omega)$ and for $C=10$ c) $t=0$ and d) $t=\pi/(2\omega)$. } \label{fig:osc1}
\end{figure}

For a time dependent pressure gradient in the horizontal direction, there is a competence between the applied external force and the gravitational field. When $\Delta P =0$, the velocity field change its direction in the neighborhood of the boundaries every half of a period. Within the material, the fluid responds with a delay in the inversion of the velocity and stagnation lines appear at the bottom and at the top of the domain (see figures \ref{fig:osc1}-(a) and (b)) due the vertical pressure gradient. When $C$ is increased, the solid matrix is harder and the fluid cannot pass easily trough the porous media without changes in its local density, the response is slower and several stagnation lines can appear as showed in figures \ref{fig:osc1}-(d) and (e). 

As the fluid goes deeper in the porous media, the amplitude of oscillation of the pressure field diminishes, when $C \gg 1$ the pressure field is approximately stationary showing just important variation in the neighborhood of the lateral walls. The numerical results for velocity field show that it has the form
\begin{equation}
{\bf u}(x,z,t) = \mathbf{\mathcal{A}}(x,z)\sin (\omega t + \phi(x,z)),\label{vel}
\end{equation} 
where $\mathbf{\mathcal{A}}$ reach its maximum at the boundaries. The phase $\phi(x,z)$ is zero at the lateral open walls and reaches its maximum within the porous media. 

The pressure of operation used was of the order of the reference pressure, and so the solid matrix deformation is small. When $C\ll 1$, the density field is approximately constant showing small spatial variations ($\Delta \rho/\rho_0 \sim   0.1\%  $), while for  values of $C$ ten times bigger variation up to $10 \%$ were found. In all the experiments, the deformation of the solid matrix was small but different from zero, showing that the pressure equation found can handle both, compressible and incompressible phenomena. 
 
\begin{figure}[h]
\begin{center}
 \includegraphics[width=1.0\columnwidth]{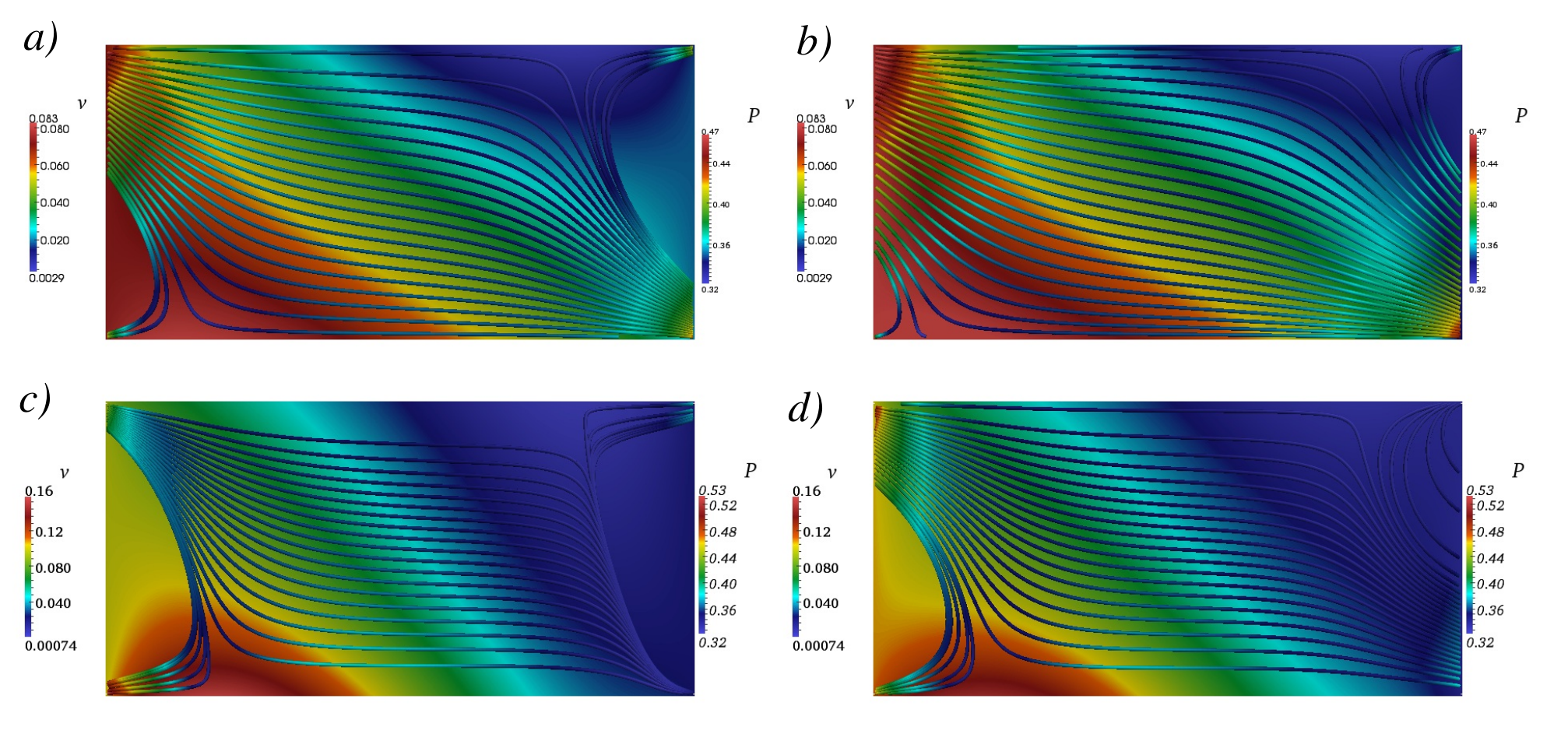} 
\end{center}
\caption{Pressure and stream lines in the plane $y=0$ for $C=0.1$ a) $t=0$ and b) $t=\pi/(2\omega)$ and for $C=10$ c) $t=0$ and d) $t=\pi/(2\omega)$. } \label{fig:osc2}
\end{figure}

When there is also a net flux due to a constant pressure gradient, $\Delta P_0 \neq 0$, the velocity field became almost independent of time inside the solid matrix, which means that the pressure field is also approximately stationary and dominated by the constant gradient $\Delta P_0$ and the gravitational field. Depending on the relative size of both, the fluid can go trough the porous media or be redirected back to the entrance, as can be appreciated in figure \ref{fig:osc2}. As the solid matrix became harder $C \gg 1$, the fluid seems to be reflected from both lateral walls as shown in figures \ref{fig:osc2}-(c) and (d).

The velocity field in this case has the same form as expression (\ref{vel}) plus a constant that depends on $\Delta P_0$. The stagnation lines observed in absence of $\Delta P_0$ persist in particular for $C\gg 1$ and the recirculation observed at the lateral walls is bigger. The mass flow through the porous media is very small, almost all the in-flow at the entrance is reflected; the fluid is heavier at the bottom due to an increased local density and the medium is harder.   

The previous two examples shows the effect of the compressibilities of the fluid and the solid matrix in time dependent problems. The Darcy law was used to show that full-time dependent problems can be modeled for incompressible or compressible flows, depending on the equations of state used and the external forcing; as in the case of oscillating flows~\cite{zhu2016}. In the present work, even when the forcing came from the boundaries, the pressure field evolves in time inside the porous media due to other effects different from inertial terms in the Darcy law.

\section{Conclusions}
\label{con}  
Starting from an equation that states that the fluid occupies all the volume available within the pores at any instant of time, we found an equation that relates the changes in the fluid volumes elements and the pressure time derivative. Hence, the problem of a single-phase flow consist of two equations for the evolution of the solid matrix and the Darcyan velocity, and a transport equation for the pressure field; the problem is closed with state equations. 

Some examples were solved numerically for systems with different values of their compressibilities, when they where assumed constant. The central result is that the time dependent variations of the pressure field can be due to several effect besides the compressibility in the flow. Moreover, the equation for the pressure found can be approximated by a non-linear diffusion equation, neglecting the velocity of deformation of the solid matrix but keeping small spatio-temporal variations in the porosity.   

The explicit dependence of the compressibilities on the effective compressibility $\mathcal{C}(\phi,P)$ permits to simulate flows in porous media with different properties. For example, the flow in aquifers where the flow and the solid matrix are incompressible or the flow through a liver in a living being, where the solid matrix can be modeled by a deformable elastic solid. 

\begin{acknowledgements}
Partial support from project UNAM-PAPIIT-IN115216 is acknowledged.
\end{acknowledgements}


\end{document}